**Strain-enhanced superconductivity in $RuO_2$ thin films: Joint experimental and theoretical studies**

Moeen ud din[1,2], Tsun-Sung Kuo[1], Yun-Chen Liao[1], Akhilesh Kr. Singh[3], Uddipta Kar[3], Wei-Li Lee[3,*], Guang-Yu Guo[1,4 *]

[1]Department of Physics and Center for Theoretical Physics, National Taiwan University, Taipei 10617, Taiwan, Republic of China

[2]Nano Science and Technology Program, Taiwan International Graduate Program, Academia Sinica, Taipei 11529, Taiwan, Republic of China

[3]Institute of Physics, Academia Sinica, Nankang, Taipei 11529, Taiwan, Republic of China

[4]Physics Division, National Center for Theoretical Sciences, Taipei 10617, Taiwan, Republic of China

We investigate the superconducting properties of $RuO_2$ films grown on $TiO_2$ substrates by combining film fabrication and transport measurement with *ab initio* calculation. Experimentally, we grow several $RuO_2$ (110) thin films on $TiO_2$ (110) substrates and observe superconductivity with critical temperature ($T_c$) of ~2.0 K in 20-nm and 30-nm $RuO_2$ (110) thin films. Theoretically, we study the superconductivity in both $RuO_2$ (110) and $RuO_2$ (100) epitaxial films within McMillan formalism of isotropic Migdal-Eliashberg theory, using the normal state properties from *ab initio* calculations. The crystal structures of the films are obtained with the out-of-plane lattice constant and atomic positions determined by *ab initio* calculations while the in-plane lattice constants held fixed to the experimentally values. Among the salient results, the calculated phonon dispersion relations of all the considered $RuO_2$ films have no imaginary frequencies, i.e., the epitaxial films are dynamically stable. Importantly, our *ab initio* calculations reveal that independent of the film orientation, the [001] compressive strain $\epsilon_{[001]}$ pushes the flat Ru $d_{z^2}$-dominant bands towards the Fermi level ($E_F$), and significantly soften the out-of-plane acoustic phonon mode near the R point in the Brillouin zone, which substantially enhances the electron-phonon coupling (EPC) strength $\lambda$ and hence the calculated $T_c$ in the $RuO_2$ films. Remarkably, the amount of the phonon softening and the enhanced $T_c$ increase almost linearly with the increasing $\epsilon_{[001]}$. The calculated $T_c$ for the films with thickness of ≥ 20 nm agree well with the experimental $T_c$. For example, the theoretical $T_c$ of 2.8 K for 20-nm $RuO_2$ (110) film and 0.8 K for 28-nm $RuO_2$ (100) film are close to the experimental $T_c$ of 2.0 K and 1.0 K, respectively. These findings demonstrate that among other things, lattice mismatch strain provides an effective route for tuning phonon-mediated superconductivity in $RuO_2$ and other thin films.

## I. INTRODUCTION

Superconductivity in transition-metal oxides (TMOs) continues to be an active area of research because of the wide range of pairing mechanisms and electronic behaviors. TMOs such as ruthenates and cuprates reveal how strong electron correlations give rise to unconventional superconductivity [1]. In particular, oxides based on ruthenium have long been a rich source of emergent quantum phenomena. Unconventional superconductivity in $Sr_2RuO_4$, metamagnetic behavior in $Sr_3Ru_2O_7$, and antiferromagnetic and ferromagnetic states in ruthenates based on calcium and strontium are notable examples [2-4]. The understanding of correlated electron systems has improved extensively as a result of these materials.

It is commonly known that rutile-structured $RuO_2$ is a highly conducting binary oxide [5-7] with numerous engineering applications, including catalysts, thin films, electrodes, and thermometers. Thin films, primarily polycrystalline films, have been produced for these applications using a variety of growth techniques [8-17]. A further and particularly striking development was the observation that superconductivity can be induced in $RuO_2$ through epitaxial strain [18,19]. While bulk $RuO_2$ remains non-superconducting down to 0.3 K [7], superconductivity with $T_c$ on the order of ~1.0 K was realized in $RuO_2$ (110) thin films grown on $TiO_2$ (110) substrates. In contrast, films grown on other rutile orientations and on alternative substrates such as $MgF_2$ and $\alpha - Al_2O_3$ (0001) and $\alpha - Al_2O_3$ $(11\bar{0}2)$ were reported to remain non-superconducting [20,21]. These results establish superconductivity in $RuO_2$ as highly sensitive to epitaxial orientation and strain anisotropy.

The microscopic origin of the superconductivity in the $RuO_2$ (110) thin films has been linked to symmetry breaking of the rutile lattice and the resulting modification of the electronic structure [22]. Epitaxial strain alters Ru-O hybridization, redistributes Ru *d*-orbital occupancies, and enhances the density of states near the Fermi level, all of which can favor superconductivity [18]. Very recently, superconductivity in strained $RuO_2$ (100) thin films was also reported, indicating that superconductivity in $RuO_2$ is not limited to the (110) orientation grown on $TiO_2$ (110) single-crystal substrates [23].

Previous theoretical and experimental studies suggested that superconductivity in $RuO_2$ thin films can be strongly influenced by epitaxial strain [18,19,23]. In particular, previous *ab initio* studies $RuO_2$ films demonstrated that strain-induced structural distortion modifies the low-energy electronic states and lattice dynamics in $RuO_2$ films, and also indicated that strain can significantly alter the Ru-O bonding environment and the associated electron-phonon interactions that govern superconductivity in $RuO_2$ films [18,19,23]. However, these theoretical works on $RuO_2$ thin films also reported the appearance of imaginary phonon modes [18,19,23], implying lattice instabilities of the strained films and thus hindering a proper description of phonon-mediated superconductivity. This raises an important question regarding the stability of strained $RuO_2$ thin films and the microscopic mechanism responsible for the emergence and enhancement of superconductivity in the $RuO_2$ films.

In this work, we combine both transport experiments and *ab initio* calculations to examine how lattice mismatch induced-strain affect the electronic structure, phonon dispersion, electron-phonon coupling, and superconductivity in $RuO_2$ thin films. Importantly, in contrast to previous theoretical predictions [18,19,23], our calculated phonon dispersions in the $RuO_2$ films do not show any imaginary frequencies, indicating that the strained $RuO_2$ films are dynamically stable. This enable us to provide a proper description of lattice dynamics, electron-phonon coupling and superconductivity in the considered $RuO_2$ films. Indeed, our calculated $T_c$ of 2.8 K for 20-nm $RuO_2$ (110) film and 0.8 K for 28-nm $RuO_2$ (100) film agree well with the experimental $T_c$ of 2.0 K from this work and 1.0 K [23], respectively.

The rest of this paper is organized as follows. In the next section, we report the results of our experiments on epitaxial $RuO_2$ (110) films on $TiO_2$ (110) substrates. In Sec. III, a brief description of the crystal structure of bulk $RuO_2$ and lattice mismatch strains in the considered $RuO_2$ films is provided. In Sec. IV, we summarize the McMillan formalism of the isotropic Migdal-Eliashberg theory of phonon-mediated superconductivity and details of our *ab initio* calculations. The calculated electronic band structure, phonon dispersion, electron-phonon coupling and superconductivity are presented and compared with available experiments in Sec. V and the supplemental material. In Sec. VI, we discuss the thickness dependence of $T_c$ in ultrathin $RuO_2$ films and also examine the effect of screened Coulomb pseudopotential $\mu^*$ on $T_c$. Finally, the conclusions drawn from this work are summarized in Sec. VII.

## II. FILM FABRICATION AND MEASUREMENT

Figure 1(a) shows the X-ray (XRD) scans of $RuO_2$ films on $TiO_2$ (110) substrates ($RuO_2(t)/TiO_2$ (110)) with nominal thicknesses ($t$) of 3-nm, 8-nm, and 20-nm, grown by using ozone-assisted molecular beam epitaxy (MBE). The reflections from the $RuO_2$ (220) and (330) planes are accompanied by notable thickness fringes, justifying the exceptional crystallinity and thickness uniformity in the $RuO_2$ thin films. By fitting to the XRD, the lattice constants of $d_{(1\text{-}10)}$, $d_{(001)}$, and $d_{(110)}$ are extracted and listed in Table I, exhibiting $t$-dependent anisotropic strain in the $RuO_2$ films as compared to the bulk $RuO_2$ lattice. For $t$ = 3 nm, the $RuO_2$ film is fully strained to the $TiO_2$ (110) substrate, resulting in a compressive strain of -4.7% along [001] and a tensile strain of about +2.3% along [1-10]. As $t$ increases to 20 nm, the compressive strain partly relaxes to about -3.1% along [001]. We also note that the film thickness extracted from the thickness fringes and low-angle reflectivity is consistent with the nominal thickness.

Table I: The extracted lattice constants of $d_{(1-10)}$, $d_{(001)}$ and $d_{(110)}$ in $RuO_2$($t$ nm)/$TiO_2$(110) with $t$ = 3 nm, 8 nm, 20 nm and 30 nm. Bulk $RuO_2$ and $TiO_2$ lattice constants are included for comparison.

| System | $d_{(110)}$ (Å) | $d_{(1-10)}$ (Å) | $d_{(001)}$ (Å) |
|---|---|---|---|
| Bulk $RuO_2$ | 3.176 | 3.176 | 3.106 |
| 30-nm film | - | 3.248 | 3.033 |
| 20-nm film | 3.230 | 3.245 | 3.010 |
| 8-nm film | 3.237 | 3.249 | 3.003 |
| 3-nm film | 3.236 | 3.248 | 2.965 |
| Bulk $TiO_2$ | 3.248 | 3.248 | 2.959 |

The temperature dependence of normalized resistivity $\rho/\rho_{3K}$ of $RuO_2$ ($t$ nm)/$TiO_2$ (110) with various thicknesses down to 1.5 K, are shown in Fig.1(b), highlighting the onset of a superconducting transition at around $T_c \approx 2.2$ K for $t$ = 20 nm and 30 nm $RuO_2$ thin films. For $t$ = 20 nm, the residual resistivity (RR) at $T$ = 5 K and the residual resistivity ratio (RRR $\equiv \rho_{300K}/\rho_{5K}$) are ~17.2 μΩ-cm and 4.0, respectively. It gives a superconducting transition temperature $T_c \approx 1.97$ K, which was defined as the temperature at which the resistivity drops by half of its normal state value. The normal state resistivity $\rho_{3K}$ monotonically increases with decreasing $t$, but the $T_c$ attains a maximum at a thickness ranging from 20 to 30 nm as shown in the inset of Fig. 2 (b). The further strain relaxation along [001] for $t$ > 30 nm and the increasing disorder for $t$ < 20 nm tend to suppress the superconductivity, resulting in a non-monotonic $t$-dependence of $T_c$ as a signature for the strain-stabilized superconductivity in $RuO_2$ thin films.

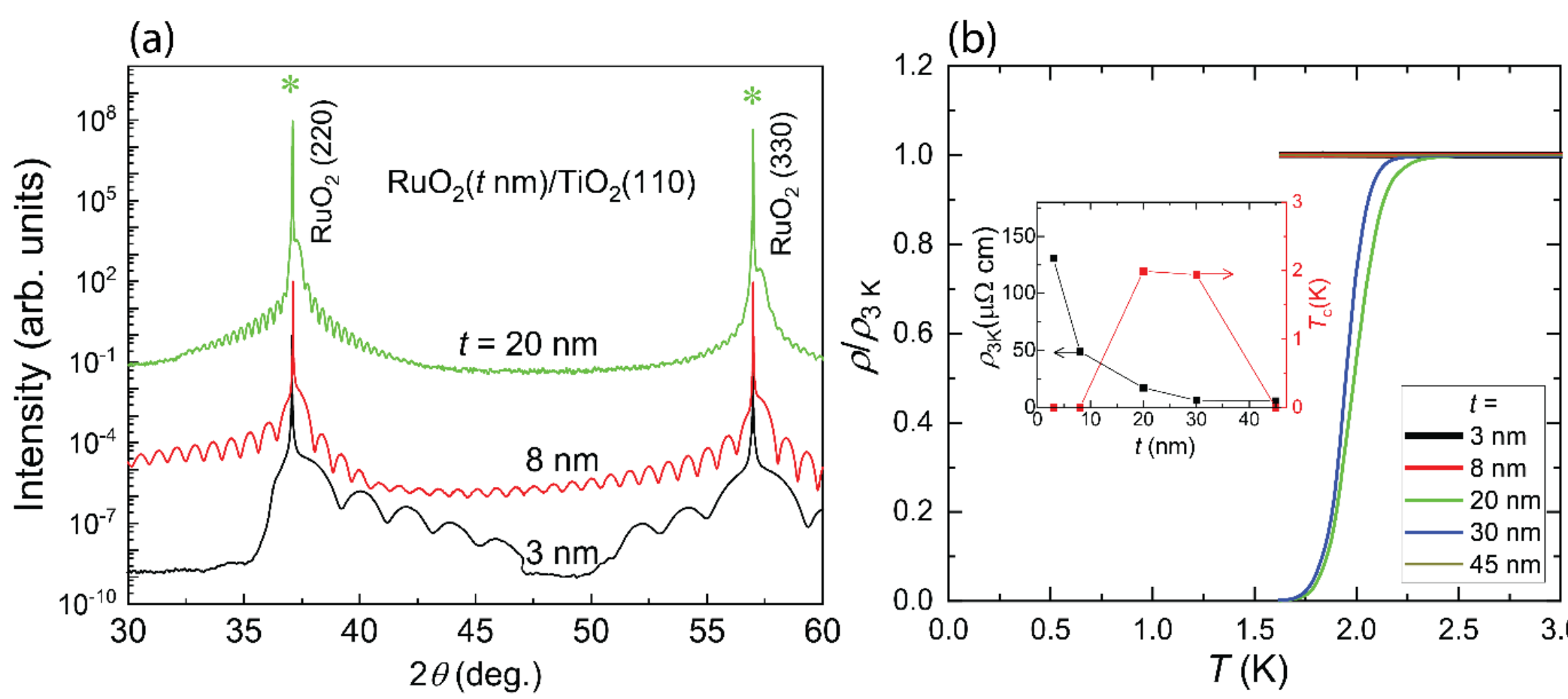

FIG. 1. XRD and transport data of $RuO_2$ ($t$ nm)/$TiO_2$(110): (a) XRD scans of $RuO_2$($t$ nm)/$TiO_2$(110) with $t$ = 3 nm, 8 nm, and 20 nm. The reflections from $TiO_2$ (110) substrate are labeled with * symbols. Pronounced thickness fringes appear around the $RuO_2$ Bragg peaks for all $t$ values, justifying the high crystallinity and uniform thickness in the $RuO_2$ films. (b) The normalized resistivity $\rho/\rho_{3K}$ versus temperature for $T \leq 3$ K in $RuO_2$ ($t$ nm)/$TiO_2$(110) with $t$ ranging from 3 nm to 45 nm, highlighting a superconducting transition with a $T_c \approx 1.97$ K for $t$ = 20 nm and non-monotonic $t$-dependent $T_c$. The inset shows the $t$-dependence of $\rho_{3K}$ and $T_c$.

## III. CRYSTAL STRUCTURE AND LATTICE MISMATCH STRAIN

Ruthenium dioxide ($RuO_2$) has a tetragonal rutile crystal structure, with point group $D_{4h}$ and space group $P4_2/mnm$ (No. 136). When $RuO_2$ is deposited on $TiO_2$ (110) substrate, lattice mismatch strain is applied on the $RuO_2$ (110) thin-film, which compress it along [001]-direction ($a'$) but extends it along $[1\bar{1}0]$-direction ($b'$) [see Fig. 2(a)]. As a result, the lattice constant of the out-of-plane (110) direction ($c'$) would expand (see Table I). The above effect lowers the point group to $D_{2h}$ while the space group becomes $Cmmm$ (No. 65). The strained $RuO_2$ (110) films thus become orthorhombic. The lattice constants of a strained $RuO_2$ (110) film can be expressed in the form

$$\begin{pmatrix} a' \\ b' \\ c' \end{pmatrix} = \begin{pmatrix} 0 & 0 & 1+\epsilon_{[001]} \\ 1+\epsilon_{[1\bar{1}0]} & 0 & 0 \\ 0 & 1+\epsilon_{[110]} & 0 \end{pmatrix} \begin{pmatrix} \sqrt{2}a_{\text{bulk}} \\ \sqrt{2}a_{\text{bulk}} \\ c_{\text{bulk}} \end{pmatrix} \quad (1)$$

where $\epsilon_{[nml]}$ is the strain along $[nml]$-direction. Here, $a_{\text{bulk}}$ and $c_{\text{bulk}}$ are the primitive cell lattice constants of bulk $RuO_2$. Detailed information is provided in Table II.

On the other hand, when $RuO_2$ is deposited on $TiO_2$ (100) substrate, the in-plane lattice constant along [010]-direction ($a'$) expands while the one along [001]-direction ( $b'$ ) is compressed. Consequently, the lattice constant of the out-of-plane (110) direction ($c'$) expands. The strain effect changes the point group to $D_{2h}$ while the space group becomes $Pnnm$ (No. 58). The strained $RuO_2$ film is orthorhombic and their lattice constants are given by

$$\begin{pmatrix} a' \\ b' \\ c' \end{pmatrix} = \begin{pmatrix} 0 & 1+\epsilon_{[010]} & 0 \\ 0 & 0 & 1+\epsilon_{[001]} \\ 1+\epsilon_{[100]} & 0 & 0 \end{pmatrix} \begin{pmatrix} a_{\text{bulk}} \\ a_{\text{bulk}} \\ c_{\text{bulk}} \end{pmatrix}. \quad (2)$$

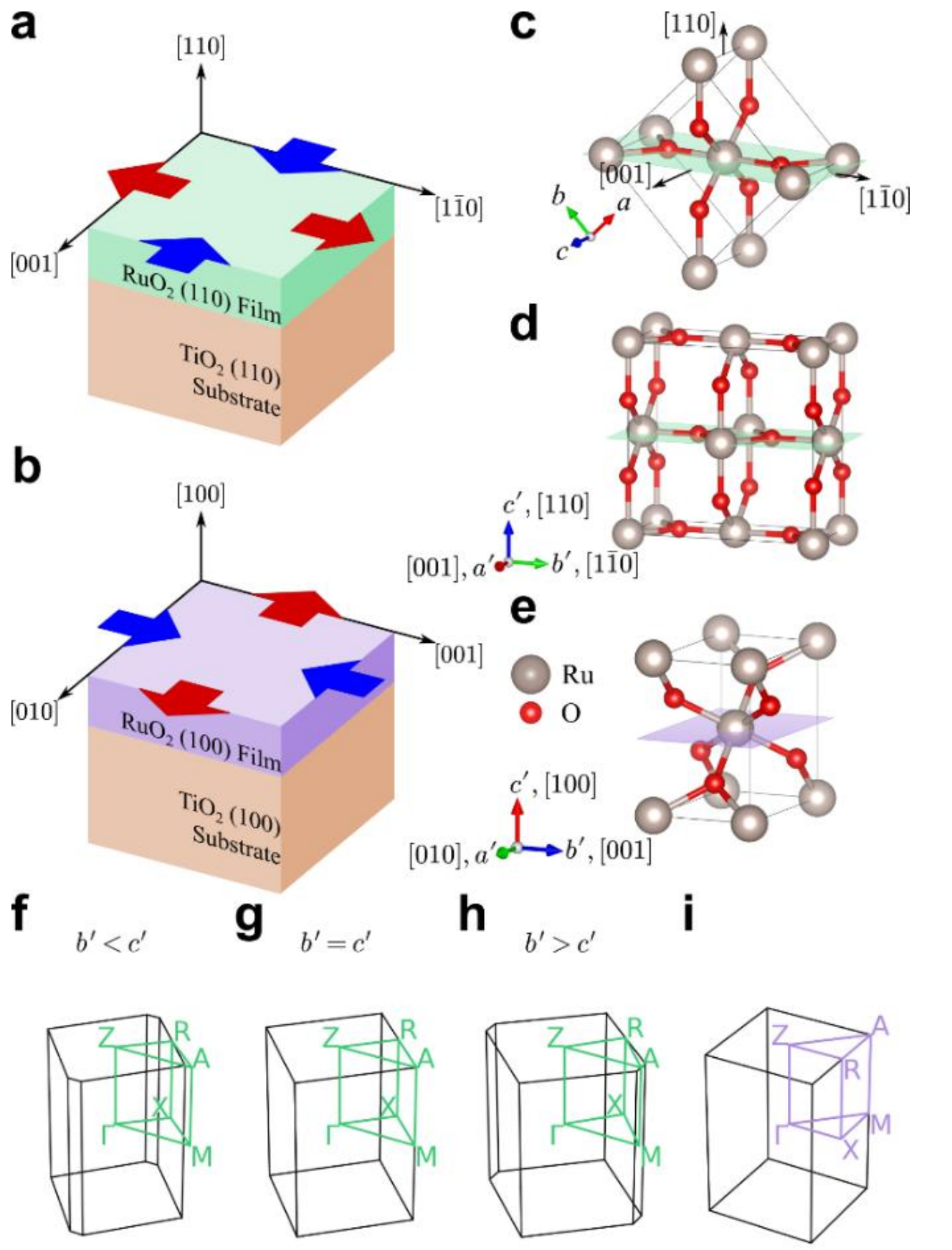

FIG. 2. (a, b) Schematic diagram of strain applied on $RuO_2$ (a) (110) thin-film and (b) (100) thin-film. (c, d, e) Crystal structure of $RuO_2$ (c) bulk, (d) (110)-oriented thin-film and (e) (100)-oriented thin-film. (f-h) Brillouin zones of (110) thin-film for (f) $b' < c'$ (corresponds to 3-nm thin-film), (g) $b' = c'$ (corresponds to bulk $RuO_2$), and (h) $b' > c'$ (corresponds to 20-nm thin-film). (i) Brillouin zone of (100) thin-film.

## IV. THEORY AND COMPUTATIONAL DETAILS

All *ab initio* structural optimizations are performed within the density functional theory

(DFT) with the generalized gradient approximation (GGA) plus the PBEsol parameterization [24,25]. The projector-augmented wave method [26], as implemented in the Vienna Ab Initio Simulation Package (VASP) [27,28], is employed. A plane-wave energy cutoff of 500 eV is adopted. Brillouin-zone integrations are carried out using the tetrahedron method with a Γ-centered 12 × 12 × 18 $k$-point mesh. Structural optimization is considered converged when the forces on all atoms are less than 0.005 eV/Å. For bulk $RuO_2$, both lattice parameters and internal atomic coordinates are fully relaxed. The calculated lattice constants of bulk $RuO_2$ are $a$ = 4.4942 Å and $c$ = 3.1097 Å, agreeing very well with the experimental values (within 0.2 %) (Table I).

$RuO_2$ (110) and (100) thin films are modeled within a strained-bulk approach (i.e., the effect of finite film thickness is neglected). This approach should work quite well for films with thicknesses of ≥ 20-nm. For ultrathin films with thicknesses of ≤ 10-nm, however, the effects of disorders, coherence length, surface and interface would become pronounced, and consequently, the *ab initio* calculations using the strained-bulk approach may not properly describe the observed superconducting properties. See Sec.VI. B for further discussion on this topic. For a $RuO_2$ film, the in-plane lattice constants $a'$ and $b'$ are fixed to the corresponding experimental values. However, the out-of-plane lattice constant ($c'$) and all internal atomic positions are allowed to relax to minimize the total energy. We note that such structural optimization calculations are needed to avoid the occurrence of the phonon modes with an imaginary frequency in the strained $RuO_2$ films [18,19,23] and hence to calculate the superconducting properties properly. The theoretically determined lattice constants $c'$ for all the $RuO_2$ (110) and (100) thin films are listed in Tables II and III, respectively. We notice that the differences between the calculated and experimental lattice constant $c'$ are less than 0.6 % (see Table I and Table II).

The electronic structure calculations are carried out using a plane-wave pseudopotential formalism, as implemented in the Quantum ESPRESSO package [29,30]. The optimized norm-conserving Vanderbilt (ONCV) scalar-relativistic pseudopotentials generated with the ONCVPSP code and the PBEsol exchange-correlation functional [31] are adopted. For ruthenium (Ru), the pseudopotential includes the $4s^2 4p^6 4d^7 5s^1$ electrons as (16) valence electrons, and for oxygen (O), it includes the $2s^2 2p^4$ electrons as (6) valence electrons. The other electrons are treated as part of the frozen cores. In all the calculations, we use plane-wave kinetic-energy and charge-density cutoffs of 90 Ry and 720 Ry, respectively. Brillouin zone integrations for the electronic structure and density of states (DOS) calculations are performed using a 16 × 16 × 24 Monkhorst-Pack $k$-point mesh. Electronic occupations are treated using Gaussian broadening of 0.05 Ry for all the considered systems except that Gaussian broadening of 0.06

Ry is used for the 3-nm $RuO_2$ (110) film. Self-consistency is achieved with an energy convergence threshold of $10^{-9}$ Ry.

Phonon dispersions, phonon DOS and electron-phonon coupling (EPC) matrix elements $g_{mn,\nu}$ (**k**,**q**) are computed using the density functional perturbation theory (DFPT) [32], as implemented in Quantum ESPRESSO [29,30]. A phonon momentum grid of 4 × 4 × 6 is used for both bulk and all $RuO_2$ thin film structures. In this work, we consider phonon-mediated superconductivity in all the considered $RuO_2$ films and bulk $RuO_2$. The superconducting transition temperatures are evaluated with the McMillan-Allen-Dynes formulas [33,34]

$$T_c = \frac{\omega_{log}}{1.2}\, exp\left[\frac{-1.04(1+\lambda)}{\lambda - \mu^*(1+0.62\lambda)}\right] \quad (3)$$

where $\omega_{log}$ is the logarithmic average phonon frequency,

$$\omega_{log} = exp\left[\frac{2}{\lambda}\int \frac{d\omega}{\omega}\alpha^2F(\omega)\, log\, \omega\right], \quad (4)$$

$\lambda$ is the dimensionless EPC constant

$$\lambda = 2\int \frac{\alpha^2F(\omega)}{\omega}d\omega, \quad (5)$$

$\mu^*$ is the averaged screened electron-electron interaction and $\alpha^2F(\omega)$ is the isotropic Eliashberg spectral function, which describes frequency-dependent EPC strength. Within the isotropic Migdal-Eliashberg theory [35],

$$\alpha^2F(\omega) = \frac{1}{2}\sum_{\nu\mathbf{q}} \omega_{q\nu}\lambda_{\mathbf{q}\nu}\delta(\omega - \omega_{\nu\mathbf{q}}), \quad (6)$$

where $\omega_{q\nu}$ is the phonon frequency of mode $\nu$ at wave vector **q**, and $\lambda_{\mathbf{q}\nu}$ is the phonon mode resolved EPC and is given by

$$\lambda_{\mathbf{q}\nu} = \frac{1}{N(E_F)\omega_{\mathbf{q}\nu}}\sum_{mn,\mathbf{k}} \left|g_{mn,\nu}(\mathbf{k},\mathbf{q})\right|^2 \delta(\varepsilon_{n\mathbf{k}} - \varepsilon_F)\delta(\varepsilon_{m\mathbf{k+q}} - \varepsilon_F). \quad (7)$$

Here $N(E_F)$ is the electronic DOS at the Fermi level $(E_F)$. Therefore, the $\lambda$ and $\omega_{log}$ can be obtained from the *ab initio* calculations. Nevertheless, $\mu^*$ is generally treated as a semiempirical parameter, and it typically varies from 0.05 to 0.30 [36-38]. Here we set $\mu^*$= 0.25 (See Sec. VI. A for the discussion on the choice of $\mu^*$).

TABLE II. Theoretical lattice parameters of bulk $RuO_2$ and strained $RuO_2$ (110) thin-films. Note that in-plane lattice constants $a'$ and $b'$ of strained $RuO_2$ (110) thin-films are taken from Table I and are fixed. $a' = d_{(001)}$ , $b' = 2d_{(1\text{-}10)}$ and $c' = 2d_{(110)}$. Lattice mismatch strains $\epsilon_{[nml]}$ are also listed, which are defined as $[(d_{(nml)}(\text{film})- d_{(nml)}(\text{bulk}))/d_{(nml)}(\text{bulk})]$.

| System | $a'$ (Å) | $\epsilon_{[001]}$ | $b'$ (Å) | $\epsilon_{[1\bar{1}0]}$ | $c'$ (Å) | $\epsilon_{[110]}$ |
|---|---|---|---|---|---|---|
| Bulk $RuO_2$ | 3.1097 | | 6.3558 | | 6.3558 | |
| 30-nm Film | 3.033 | -2.466% | 6.495 | 2.198% | 6.4323 | 1.204% |
| 20-nm Film | 3.010 | -3.206% | 6.490 | 2.180% | 6.5100 | 1.591% |
| 8-nm Film | 3.003 | -3.431% | 6.498 | 2.175% | 6.4650 | 1.681% |
| 3-nm Film | 2.965 | -4.653% | 6. 496 | 2.171% | 6.4590 | 2.418% |

TABLE III. Theoretical lattice parameters of bulk $RuO_2$ and strained $RuO_2$ (100) thin-films. Note that the experimental in-plane lattice constants $a'$ and $b'$ of strained $RuO_2$ (100) thin-films from [23] are used. Lattice mismatch strains $\epsilon_{[nml]}$ are also listed, which are defined as $[(d_{(nml)}(\text{film})- d_{(nml)}(\text{bulk}))/d_{(nml)}(\text{bulk})]$.

| System | $a'$ (Å) | $\epsilon_{[010]}$ | $b'$ (Å) | $\epsilon_{[001]}$ | $c'$ (Å) | $\epsilon_{[100]}$ |
|---|---|---|---|---|---|---|
| Bulk $RuO_2$ | 4.494 | | 3.110 | | 4.494 | |
| 28-nm Film | 4.580 | 1.914 % | 3.040 | -2.241 % | 4.518 | 0.534% |
| 17-nm Film | 4.580 | 1.914 % | 3.020 | -2.885 % | 4.543 | 1.090% |

## V. RESULTS

### A. Electronic band structure

In Fig. 3, we present the calculated electronic band structures along high-symmetry paths in the Brillouin zone and also DOS for bulk $RuO_2$ as well as strained 20-nm $RuO_2$ (110) and 28-nm $RuO_2$ (100) films. The corresponding orbital-projected band structures and atom-decomposed DOS spectra are displayed in Fig. S1 in the Supplementary Material (SM) [39]. The calculated band structures and DOS spectra for strained 30-nm, 8-nm, and 3-nm $RuO_2$ (110), and 17-nm $RuO_2$ (100) films are presented in Fig. S2 in SM [39]. The crystal field splits Ru *d*-orbital dominant bands to Ru $e_g$ bands and $t_{2g}$ bands. Since we focus on the superconducting properties of the considered $RuO_2$ systems in this work, we plot the energy bands and DOS spectra only in the energy range from 2.0 eV below $E_F$ to 2.0 eV above $E_F$. All the energy bands in this energy window are Ru $t_{2g}$ bands [40]. Ru $e_g$ orbitals hybridize strongly with 12 O *p* orbitals, and consequently, twelve O *p*-dominated bonding bands lie below -2.0 eV and four Ru $e_g$-dominated antibonding bands lie above 2.0 eV [40]. Because of the coordinate systems used in the present calculations, Ru $t_{2g}$-dominated bands include all five *d*-orbitals, namely, $d_{xy}$, $d_{xz}$, $d_{yz}$, $d_{x^2-y^2}$ and $d_{z^2}$.

Figure 3, Figs. S1 and S2 show that in both bulk $RuO_2$ and all considered $RuO_2$ films, four energy bands cross the Fermi level. This confirms their metallic behavior, as observed in the present and many previous experiments [5-7,18,19]. Figure S1 in SM [39] further indicates that the energy bands that cross $E_F$ arise from Ru $d_{xy}$, $d_{xz}$ orbitals along Γ-Z direction, and $d_{x^2-y^2}$ orbitals along Γ-X and M-Γ paths. One notable feature in the band structures of all the considered $RuO_2$ systems is the occurrence of two flat Ru $d_{z^2}$ dominated bands along the in-plane Z-R-A-Z directions (see Fig. 3, Figs. S1 and S2), mainly due to the weak interactions among Ru $d_{z^2}$ orbitals in the in-plane directions and being consistent with the isotropic nature of the rutile lattice. Interestingly, there are also two flat bands of $d_{xy}$, $d_{xz}$ and $d_{xy}$ characters along the M-Γ direction.

Figure 3, Figs. S1 and S2 indicate that the calculated energy bands in all the considered $RuO_2$ films are similar to that of bulk $RuO_2$. Furthermore, these figures also suggest that the DOS spectra of the $RuO_2$ films are qualitatively the same as that of bulk $RuO_2$. All the considered $RuO_2$ systems show a finite DOS at the Fermi level, consistent with their observed metallic behavior. Moreover, all the considered systems exhibit a pronounced peak in the DOS spectrum slightly below $E_F$, due to the prominent flat Ru $d_{z^2}$ dominated bands along the in-plane Z-R-A-Z directions noted above.

However, noticeable modifications caused by the biaxial strains in the considered $RuO_2$ films are observed in the vicinity of the Fermi level. In particular, Fig. 3 and Fig. S2 in the SM [39] show that the flat Ru $d_{z^2}$ bands along the Z-R-A-Z directions move towards $E_F$ monotonically with increasing compressive strain $\epsilon_{[001]}$ due to the decreasing film thickness. Consequently, the prominent peak in the DOS spectrum also shifts toward $E_F$. Nevertheless, they remain about 0.5 eV below $E_F$ (see Fig. 3, Fig. S1 and Fig. S2), and thus the shift hardly modifies the DOS at $E_F$ (see Table V). The shift of these flat bands is observed in both $RuO_2$ (110) and $RuO_2$ (100) films, being nearly orientation-independent (see Fig. 3). For example, Fig. 3 shows that the shift of the flat bands is more pronounced in the 20-nm $RuO_2$ (110) film than in the 28-nm $RuO_2$ (100) film. Interestingly, this could be correlated with the relative calculated and experimental $T_c$ values in these $RuO_2$ films (see Table V). Nevertheless, the increase in $N(E_F)$ is rather insignificant (see Table V).

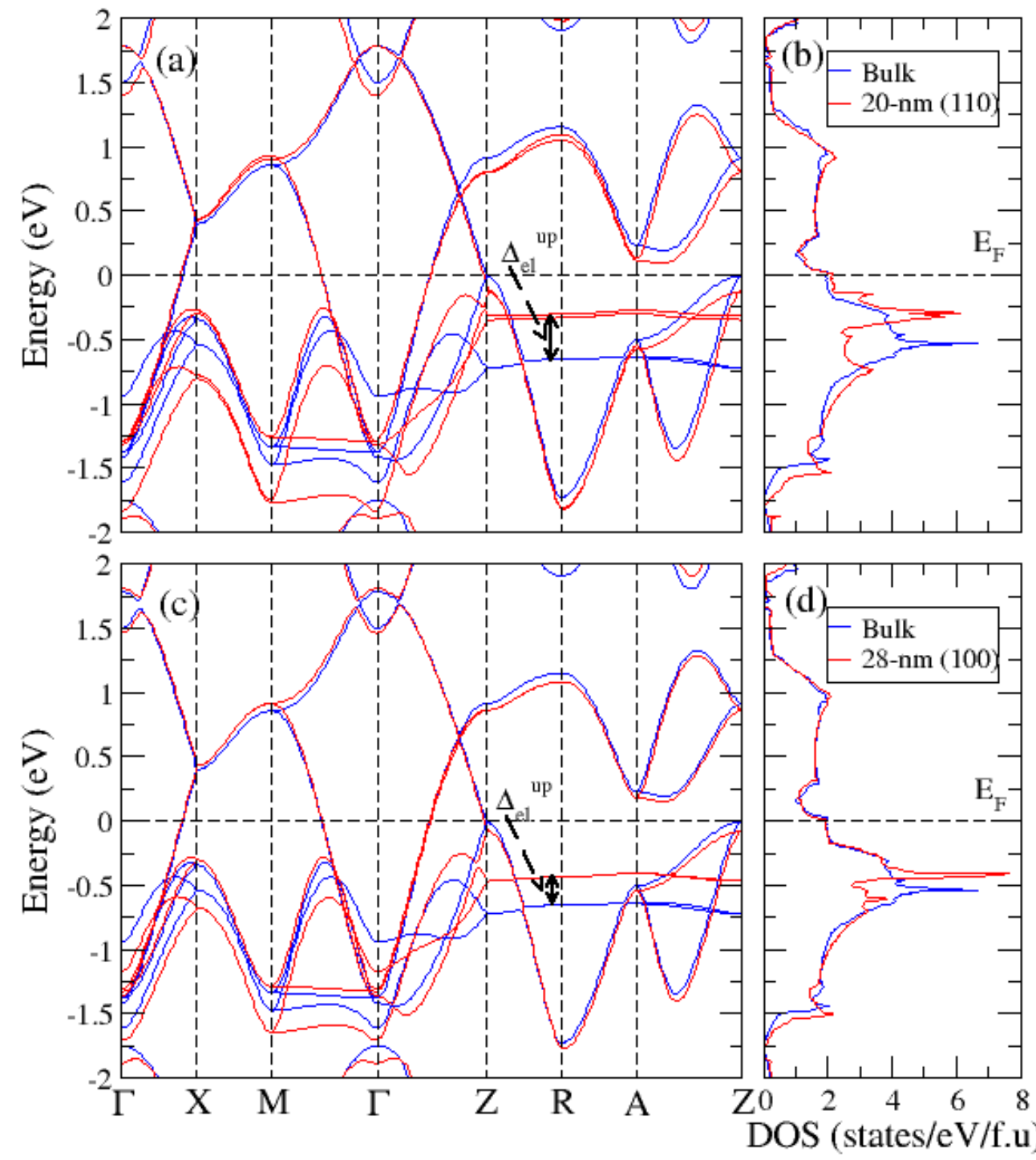

FIG. 3. (a, c) Electronic band structure and (b, d) density of states (DOS) for bulk $RuO_2$, 20-nm $RuO_2$ (110) (b), and 28-nm $RuO_2$ (100) (d) thin films. The double arrowed lines denote the upwards shift $\Delta_{el}^{up}$ of the flat Ru $d_{z^2}$ bands near R due to the compressive strain $\epsilon_{[001]}$ in the $RuO_2$ films.

### B. Phonon dispersion relations

The calculated phonon dispersion, total and atom-projected phonon DOS spectra of bulk $RuO_2$ are depicted in Fig. 4. Because the primitive unit cell contains six atoms, there are 18 phonon bands in Fig. 4(a), three acoustic modes and 15 optical modes. The frequencies of the optical modes at the BZ center are listed in Table IV, which can be compared with optical experiments. Indeed, the frequencies of Raman active phonon modes have been measured [41]. Table IV indicates that the calculated and measured frequencies of Raman active modes agree quite well with each other, with the differences being within 8 %. Figure 4(b) shows that the PhDOS spectrum below 36 meV mainly results from the vibrations of heavy Ru atoms, while the vibrations of light O atoms dominate the DOS above 36 meV.

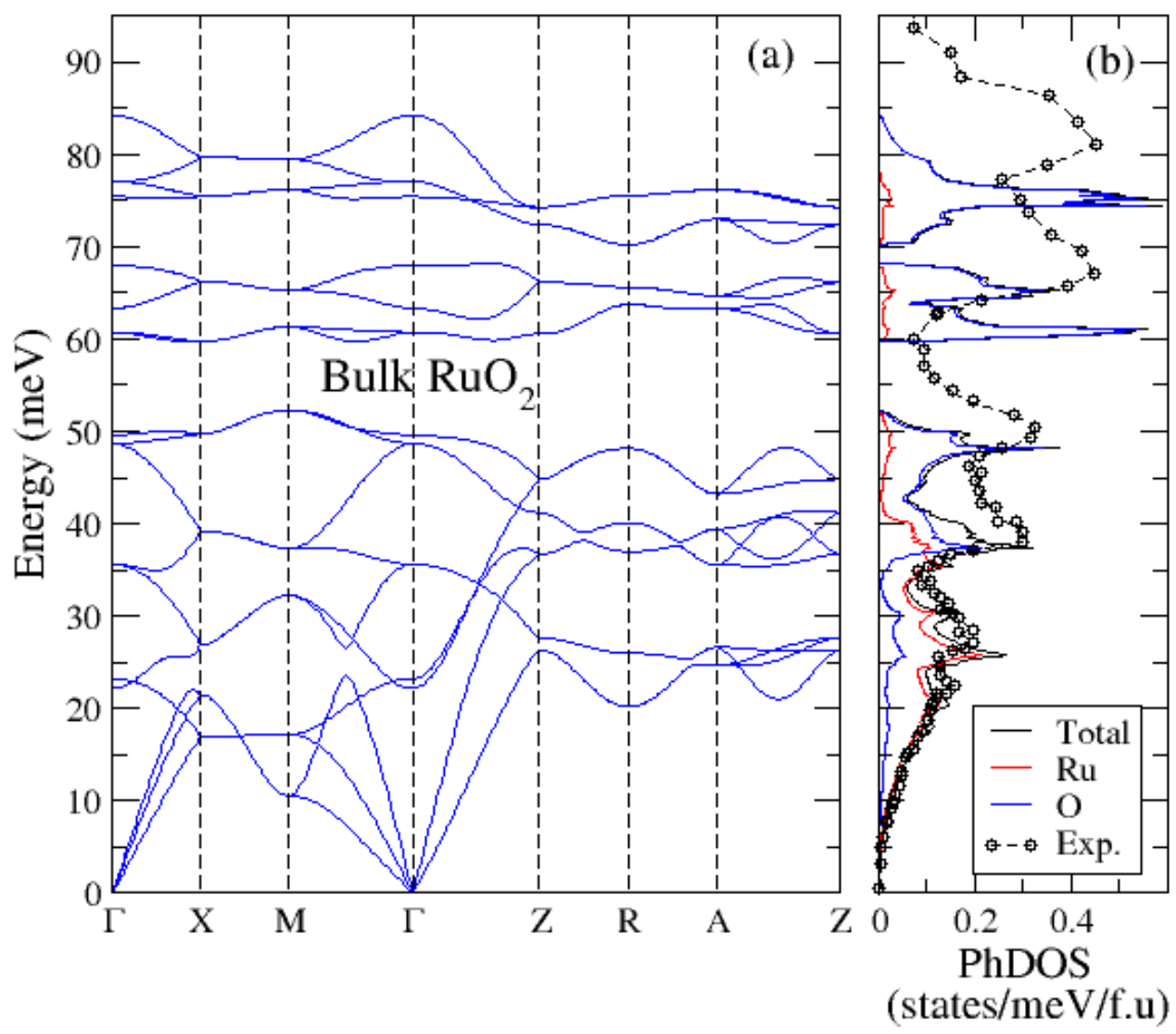

FIG. 4. (a) Phonon dispersion and (b) total and atom-projected phonon density of states (PhDOS) of bulk $RuO_2$. Experimental PhDOS from inelastic neutron scattering [42] is also plotted in (b) for comparison.

The calculated total PhDOS spectrum agrees quite well with that determined by neutron scattering experiments [42], especially for phonon energies below 50 meV. Together with the good agreement between the calculated and measured lattice constants, we conclude that our *ab initio* calculations using the GGA [24] with the PBEsol parameterization [25] can describe well the ground state structural properties and lattice vibrations in the $RuO_2$ system.

TABLE IV. Calculated frequencies and optical activities [infra-red (IR) active, Raman (R) active or optically inactive (Silent)] of the phonons at the Brillouin zone center of bulk $RuO_2$. Measured frequencies from Raman experiments [41] are also listed for comparison.

| Mode | Frequency (meV) | Activity | Frequency[a] (meV) |
|---|---|---|---|
| $B_{1g}$ | 22.2 | R | 20.5 |
| $B_{1u}$ | 23.2 | Silent | |
| $E_u$ | 35.6×2 | IR | |
| $E_u$ | 48.9×2 | IR | |
| $A_{2g}$ | 49.5 | Silent | |
| $E_g$ | 60.7×2 | R | 65.2 |
| $A_{2u}$ | 63.3 | IR | |
| $B_{1u}$ | 67.9 | Silent | |
| $A_{1g}$ | 75.4 | R | 80.0 |
| $E_u$ | 77.0×2 | IR | |
| $B_{2g}$ | 84.2 | R | 88.7 |

[a] Frequencies from Raman experiments [41].

The phonon dispersion and PhDOS in all the considered $RuO_2$ films are plotted in Fig. 5 and Fig. S3 in the SM [39]. It is clear from Fig. 5 and Fig. S3 that overall, the phonon dispersion and PhDOS in all the $RuO_2$ epitaxial films are similar to bulk $RuO_2$. For example, both phonon dispersion and PhDOS spectra in all the considered systems span approximately from 0 to slightly above 80 meV, and have a gap of about 7 meV located at 55 meV. Importantly, as for bulk $RuO_2$, the calculated phonon dispersions in all the considered $RuO_2$ films exhibit no imaginary frequencies, in contrast to all the previous *ab initio* calculations [18, 19, 23]. This demonstrates that all the considered $RuO_2$ (110) and $RuO_2$ (100) films are dynamically stable. This allows us to evaluate the superconducting properties such as $T_c$ of $RuO_2$ films more reliably than before [18, 23].

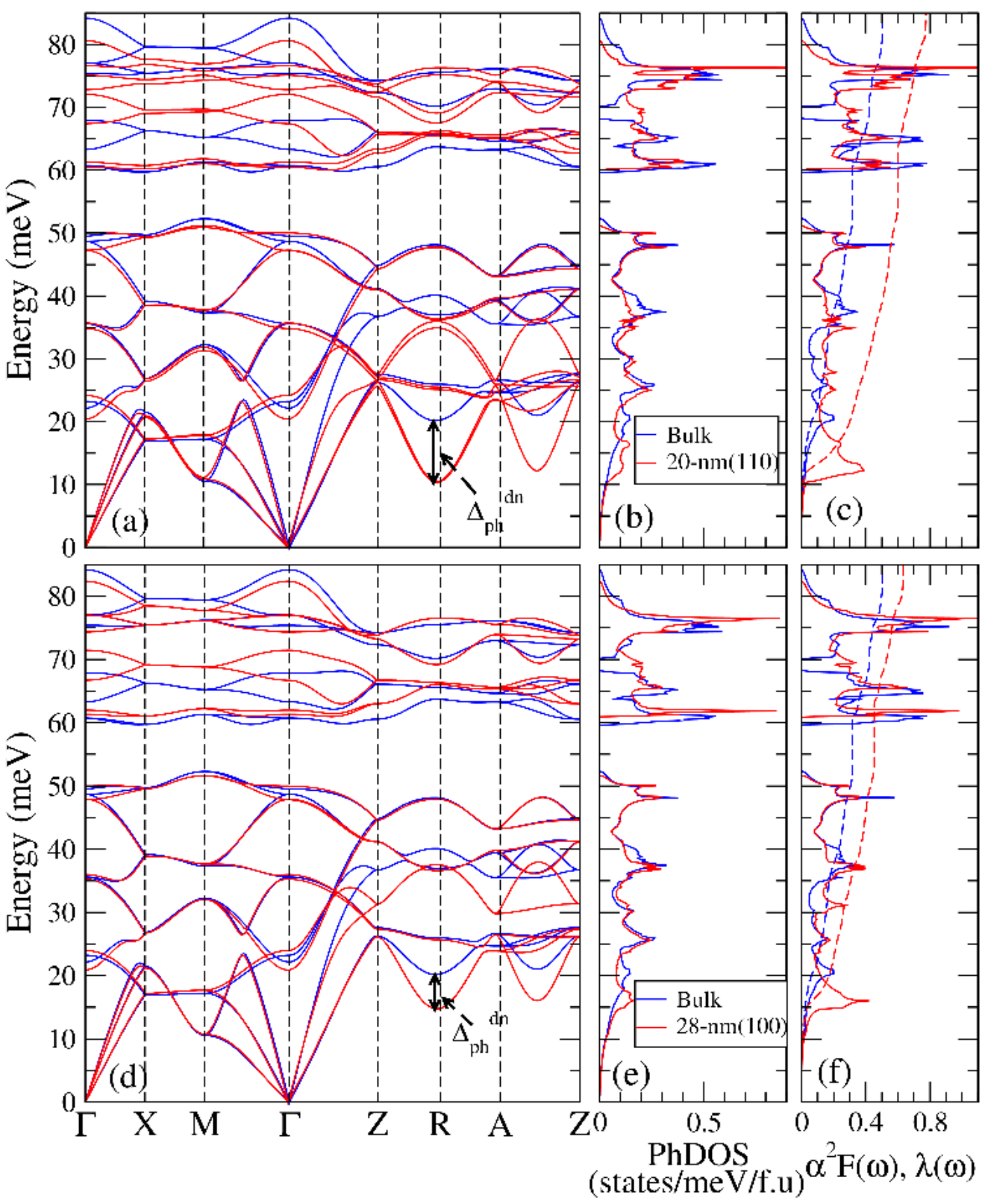

FIG. 5. (a, d) Phonon dispersions (b, e) phonon density of states (PhDOS) and (c, f) Eliashberg spectral function $\alpha^2F(\omega)$ (solid lines) and frequency-dependent electron-phonon coupling strength $\lambda(\omega)$ (dashed lines) for bulk $RuO_2$, 20-nm $RuO_2$ (110) and 28-nm $RuO_2$ (100) thin films. The double arrowed lines denote the lowering $\Delta_{ph}^{dn}$ of the lowest phonon mode frequency near R due to the compressive strain $\epsilon_{[001]}$ in the $RuO_2$ films.

However, significant changes caused by the lattice mismatch strains do exist. One pronounced change in the phonon dispersion due to the compressive strain in the $RuO_2$ films is as follows. In all considered $RuO_2$ films, the low-energy (below 40 meV) optical and out-of-plane acoustic branches along the Γ-Z-R-A lines show significant softening (see Fig. 5 and Fig. S3).

Moreover, as the compressive strain $\epsilon_{[001]}$ increases with the decreasing film thickness, this phonon softening becomes monotonically larger. In fact, as Fig. 6 shows, this phonon softening is nearly proportional to the compressive strain $\epsilon_{[001]}$. Interestingly, this is correlated with the upwards shift of the flat Ru $d_{z^2}$ bands near the R point in the BZ (Fig. 3 and Fig. S2 in the SM), which is also proportional to $\epsilon_{[001]}$. Ru $t_{2g}$ -orbitals also hybridize significantly with O $p$-orbitals, and the flat Ru $d_{z^2}$ bands arise from the antibonding bonding of the Ru $d_{z^2}$-O $p_z$ orbitals. Consequently, the energy levels of the flat Ru $d_{z^2}$ bands are sensitive to the compressive strain along the (001) direction.

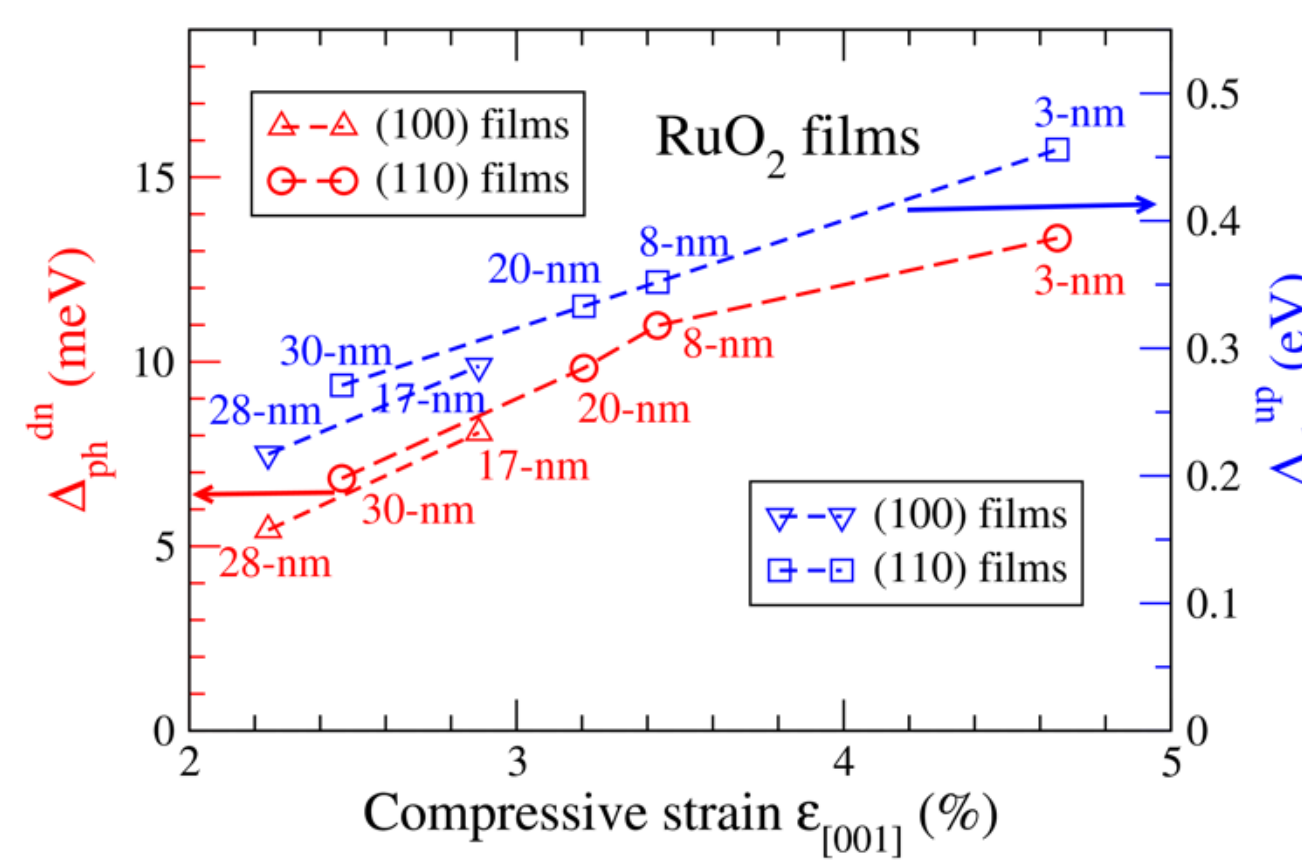

FIG. 6. The lowering $\Delta_{ph}^{dn}$ of the lowest phonon mode frequency (see Fig. 4 and Fig. S3) and the upwards shift $\Delta_{el}^{up}$ of the flat Ru $d_{z^2}$ bands (see Fig. 3 and Fig. S2) at the R point in the Brillouin zone versus the compressive strain $\epsilon_{[001]}$ in the considered $RuO_2$ films.

Importantly, these softening modes give rise to a noticeable peak at ~15 meV in the PhDOS, which leads to a prominent peak in the Eliashberg functions $\alpha^2F(\omega)$ in the low frequency region in the $RuO_2$ films [Figs. 5c) and 5f) as well as Figs. S3(c) and S3(f)]. Equation (3) indicates that this low frequency peak in the $\alpha^2F(\omega)$ will result in a significant increase in the EPC constant $\lambda$ and hence enhances the superconductivity in the $RuO_2$ films, as will be elaborated in the next section.

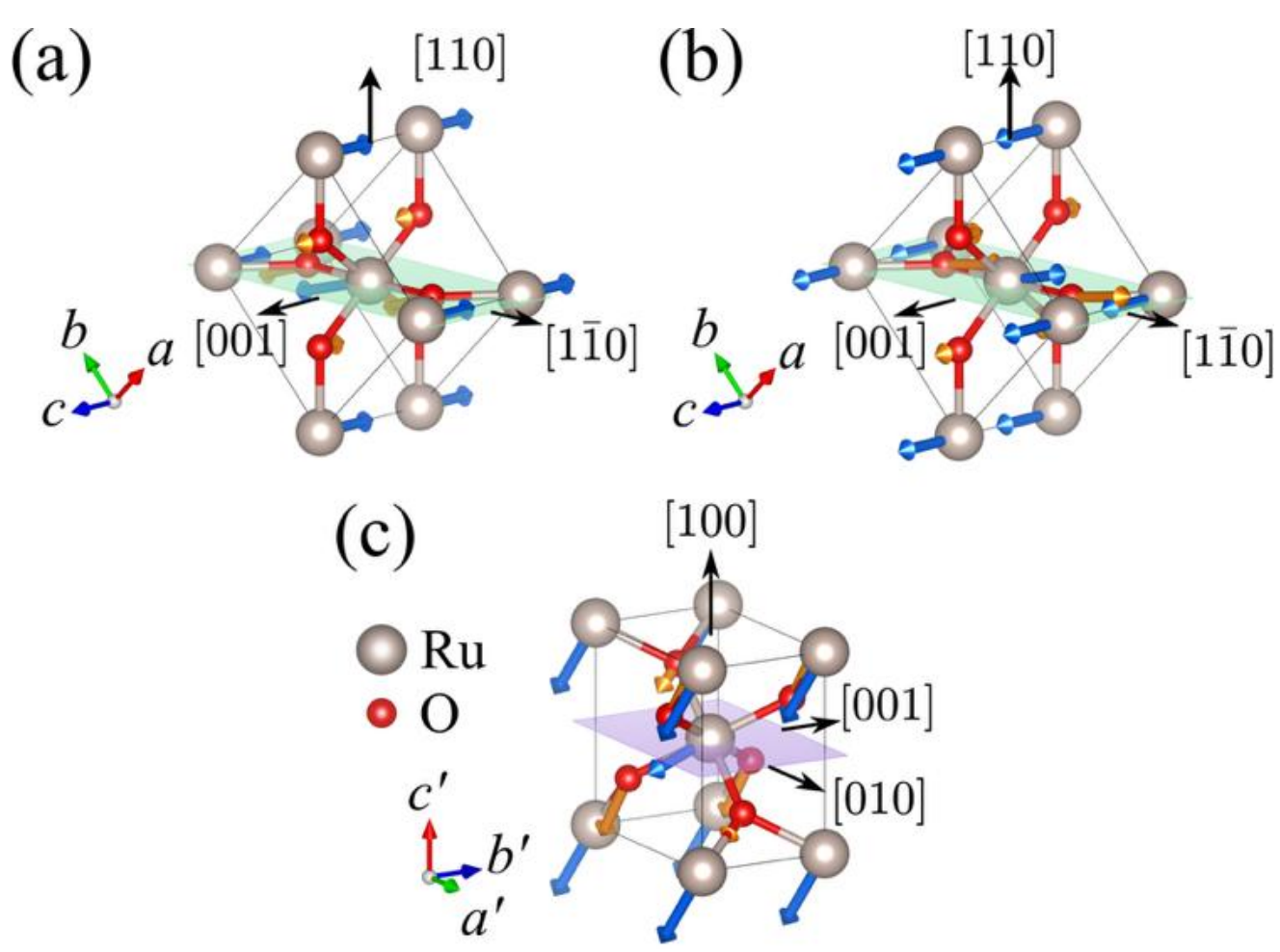

FIG. 7. Atomic displacement patterns of the first phonon mode at the R point in (a) bulk $RuO_2$, (b) 20-nm $RuO_2$ (110) film and (c) 28-nm $RuO_2$ (100) film.

To further understand the origin of this intriguing phonon softening, we plot the atomic displacement patterns of the first phonon mode at the R point in bulk $RuO_2$ as well as in strained 20-nm $RuO_2$ (110) and 28-nm $RuO_2$ (100) films in Fig. 7. We find that, for bulk $RuO_2$ and the 20-nm $RuO_2$ (110) film, the atomic vibrations consist of two in-plane components: one along the [001] direction and the other primarily along the [1$\bar{1}$0] direction, with a slight deviation from the latter [see Figs. 7(a) and 7(b)]. Similarly, for the 28-nm $RuO_2$ (100) film, the mode vibrations also have two components, one along the [001] direction but the other along the [110] direction. Thus, in both strained $RuO_2$ (110) and $RuO_2$ (100) films, the Ru and O vibrations of the first phonon mode at the R point contain the [001] component, along which the lattice is compressed by the lattice mismatch strains in the films. This phonon softening can be attributed to the antibonding nature of the Ru $d_{z^2}$-O $p_z$ states. Under compressive strain, the overlap between the Ru $d_{z^2}$ and O $p_z$ orbitals is enhanced, shifting the occupied antibonding states to higher energy (see Fig. 3) and thereby weakening the Ru–O bonding interaction. Consequently, the corresponding bond stiffness is reduced, leading to the observed phonon softening (see Figs. 5 and 6).

### C. Electron-phonon coupling and superconductivity

Figures 5(c) and 5(f) as well as Figs. S3(c) and S3(f) present the calculated Eliashberg spectral function $\alpha^2F(\omega)$ (solid lines) and cumulative frequency-dependent EPC strength $\lambda(\omega)$ (dashed lines). Table V lists the calculated EPC constant $\lambda$, logarithmic average phonon frequency $\omega_{log}$, and the superconducting transition temperature $T_c$ for bulk $RuO_2$ and all the considered $RuO_2$ films. First of all, these figures indicate that in all the considered systems, the $\alpha^2F(\omega)$ spectrum follows closely the PhDOS spectrum. Second, overall, the $\alpha^2F(\omega)$ spectra of the $RuO_2$ films are very similar to that of bulk $RuO_2$. One notable difference is the occurrence of the prominent peak at around 15 meV in all the considered $RuO_2$ films. As for the PhDOS spectra, this peak results from the softening of the lowest phonon band centered at R due to the compressive strain $\epsilon_{[001]}$ (see Fig. 5 and Fig. S3 in the SM [39]), and leads to the significantly increased EPC constant $\lambda$ for the $RuO_2$ films, as listed in Table V. In particular, the calculated EPC constant $\lambda$ for bulk $RuO_2$ is 0.50, while that for 20-nm $RuO_2$ (110) film and 28-nm $RuO_2$ (100) film are 0.78 and 0.63, respectively. This gives rise to the significantly enhanced theoretical $T_c$ of 2.77 K in 20-nm $RuO_2$ (110) film and 0.83 K in 28-nm $RuO_2$ (100) film, compared with the theoretical $T_c$ of 0.06 K in bulk $RuO_2$. We notice that the calculated $T_c$ of 2.77 K 20-nm $RuO_2$ (110) is in good agreement with the experimental $T_c$ of 1.97 K. For 28-nm $RuO_2$ (100), the calculated $T_c$ of 0.83 K also agree well with the measured $T_c$ of 0.97 K [23]. No superconductivity has been observed in bulk $RuO_2$ down to 0.3 K [7], while the calculated $T_c$ is 0.06 K. Table V shows that the DOS at $E_F$ in the strained $RuO_2$ films hardly increases, indicating that the enhanced superconductivity in the strained $RuO_2$ films originates mainly from the increased EPC constant $\lambda$ caused by the softening of the phonon modes near the R point due to the compressive strain $\epsilon_{[001]}$.

TABLE V. Calculated electron-phonon coupling constant $\lambda$, logarithmic average phonon frequency $\omega_{log}$, density of states at the Fermi level $N(E_F)$, and superconducting transition temperature $T_c$ of bulk $RuO_2$ as well as all considered $RuO_2$ (110) and $RuO_2$ (100) films. Theoretical $T_c$ are calculated with screened Coulomb repulsion $\mu^*$ = 0.25. Experimental $T_c$ values are also listed for comparison.

| System | $\epsilon_{[001]}$ (%) | $\lambda$ | $\omega_{log}$ (K) | $N(E_F)$ (states/eV/f. u) | $T_c^{The.}$ (K) | $T_c^{Exp.}$ (K) |
|---|---|---|---|---|---|---|
| Bulk $RuO_2$ | | 0.50 | 459.3 | 1.00 | 0.06 | < 0.3 [a] |
| 30-nm (110) | 2.350 | 0.67 | 363.1 | 1.04 | 1.28 | 1.97 [b] |
| 20-nm (110) | 3.097 | 0.78 | 315.7 | 1.05 | 2.77 | 1.97 [b] |
| 8-nm (110) | 3.316 | 0.82 | 297.7 | 1.05 | 3.47 | < 1.5 [b] |
| 3-nm (110) | 4.540 | 0.95 | 253.9 | 1.05 | 5.33 | < 1.5 [b] |
| 28-nm (100) | 1.914 | 0.63 | 385.9 | 1.02 | 0.83 | 0.97 [c] |
| 17-nm (100) | 1.914 | 0.71 | 344.2 | 1.03 | 1.84 | 0.94 [c] |

[a] Experimental value from Lin *et al.* [7]; [b] Experimental values from this work. [c] Experimental data from Wadehra *et al.* [23].

Figure 8 displays the calculated superconducting transition temperatures $T_c$ as a function of compressive strain $\epsilon_{[001]}$ for both $RuO_2$ (110) and (100) films together with available experimental data. For the $RuO_2$ (110) films, Fig. 8(a) indicates that the calculated $T_c$ increases almost linearly with the increasing $\epsilon_{[001]}$ strain, evolving from a low value of 1.28 K in 30-nm $RuO_2$

(110) film to a much higher value of 5.33 K in ultrathin 3-nm $RuO_2$ (110) film (see Table V). Note that the calculated $T_c$ for bulk $RuO_2$ is only 0.06 K, indicating that the compressive strain $\epsilon_{[001]}$ can cause tens folds of the enhancement of $T_c$ in the $RuO_2$ (110) films. Another notable trend is that as the thickness of the $RuO_2$ (110) film is reduced, e.g., from 30-nm to 3-nm, the enhancement of the $T_c$ becomes more pronounced. This is caused by the increased $\epsilon_{[001]}$ due to the reduced film thickness, and hence more strongly enhanced EPC constant.

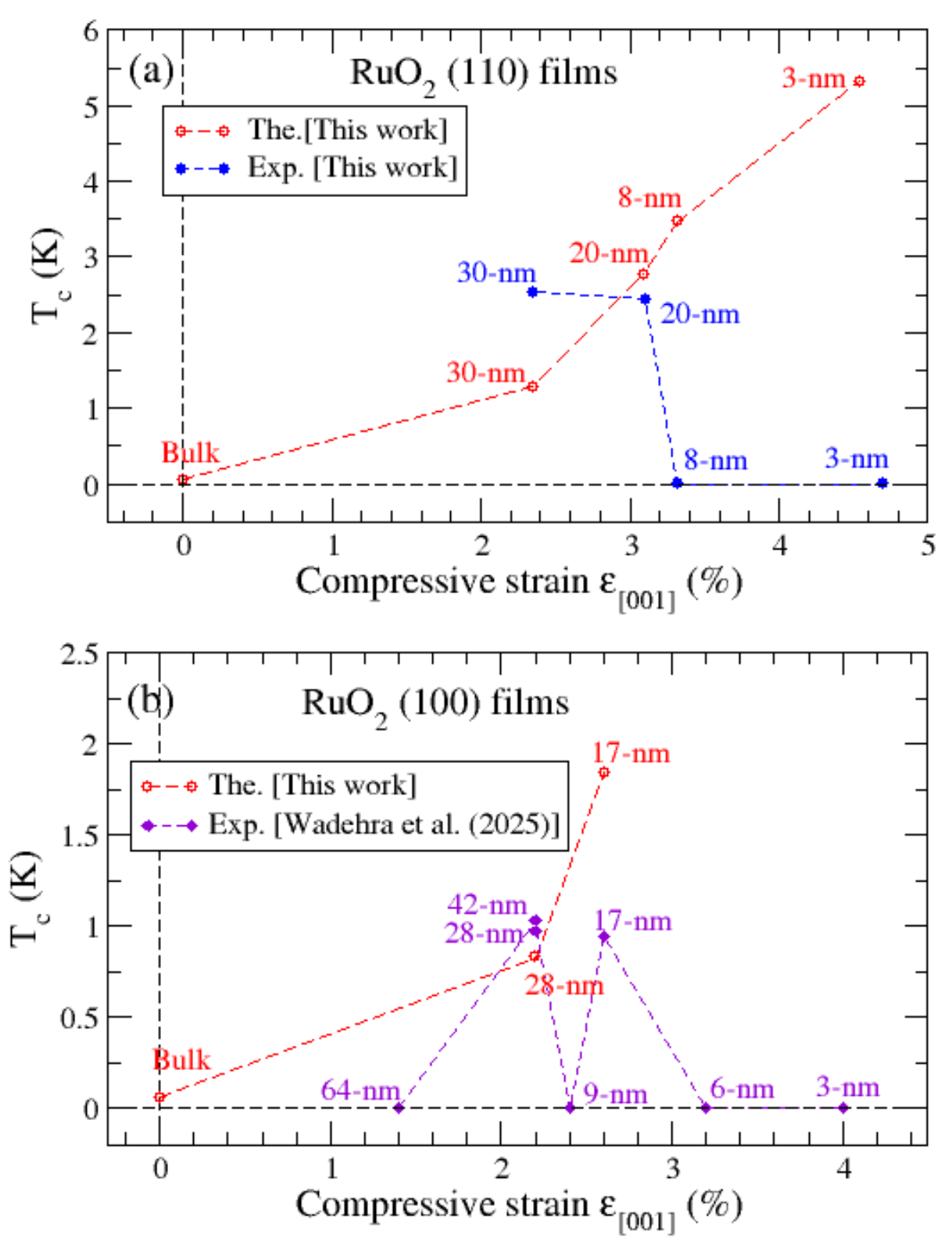

FIG. 8. Superconducting transition temperature $T_c$ versus compressive strain $\epsilon_{[001]}$ for (a) $RuO_2$ (110) films and (b) $RuO_2$ (100) films. The theoretical $T_c$ values are calculated using Eq. (3) and $\mu^* = 0.25$. The experimental data from this work and from Wadehra *et al.* [23], are also plotted for comparison.

Similarly, for the $RuO_2$ (100) films, Fig. 8(b) suggests that the calculated $T_c$ increases monotonically with the $\epsilon_{[001]}$, although only two $RuO_2$ (100) films have been studied. As mentioned in the preceding paragraphs, Fig. 8 shows that for thicker $RuO_2$ films ($t \geq$ 20 nm), the calculated $T_c$ generally agree quite well with the observed $T_c$. However, even in ultrathin $RuO_2$ films with thickness $t < 10$ nm, the calculated $T_c$ continuously increases with decreasing film thickness, consistent with the increasing compressive strain, phonon softening, and enhanced electron-phonon coupling. Nevertheless, the superconductivity is not observed experimentally in these ultrathin $RuO_2$ films (see Fig. 8, and [18,19,23]). The suppression of the superconductivity may be attributed to large disorder and strong interface scattering in these ultrathin $RuO_2$ films, as will be further discussed in Sec. VI. B below.

## VI. DISCUSSION

### A. Effect of Coulomb pseudopotential $\mu^*$ on $T_c$

To evaluate the superconducting $T_c$ using Eq. (3) with the EPC constant $\lambda$ and logarithmically averaged phonon frequency $\omega_{log}$ from the *ab initio* calculations, an appropriate choice of the semiempirical Coulomb pseudopotential $\mu^*$ is required. Equation (3) indicates that in general, a nonzero $\mu^*$ value suppresses phonon-mediated superconductivity and hence reduces $T_c$. One way to proceed is that a value of $\mu^*$ is selected such that the calculated transition temperature of a well-known material aligns with the experimental $T_c$ of the material. Therefore, we compute $T_c$ for bulk $RuO_2$, 20-nm $RuO_2$ (110) and 28-nm (100) thin films using Eq. (3) for several different $\mu^*$ values, and the results are summarized in Fig. 9. Figure 9 shows that $T_c$ decreases monotonically with increasing $\mu^*$, reflecting the progressive suppression of the phonon-mediated electron pairing interaction by the screened Coulomb repulsion. No superconductivity has been observed down to 0.3 K in bulk $RuO_2$ [7], i.e., the upper bound of $T_c$ of bulk $RuO_2$ is 0.3 K. Since the calculated $T_c$ of bulk $RuO_2$ is 0.06 K when $\mu^*$ = 0.25, and is zero when $\mu^*$ = 0.30, we set $\mu^*$ = 0.25 in this work. Furthermore, Table V shows that when $\mu^*$ = 0.25, the calculated $T_c$ for the 28-nm $RuO_2$ (100) film is 0.83 K, in good agreement with the measured value of 0.97 K (Table III). Also, the calculated $T_c$ for 20-nm $RuO_2$ (110) film is 2.77 K, being close to the experimental value of 1.97 K (Table V). This supports that $\mu^*$ = 0.25 provides a reasonable description of the superconducting properties of strained $RuO_2$ thin films.

Interestingly, Fig. 9 shows that 20-nm $RuO_2$ (110) film exhibits the highest $T_c$ over the entire range of $\mu^*$, while bulk $RuO_2$ consistently shows the lowest values. The calculated $T_c$ of 28-nm (100) film is between 20-nm $RuO_2$ (110) film and bulk $RuO_2$. This can be correlated with the fact that the compressive strain $\epsilon_{[001]}$ in 28-nm (100) film is smaller than in 20-nm (110) film, i.e., strain enhanced EPC strength is stronger in 20-nm (110) film than in 28-nm (100) film. All these results thus demonstrate that although the absolute value of $T_c$ depends on the choice of $\mu^*$, our predicted strain enhancement of superconductivity in $RuO_2$ (110) and (100) films remains robust.

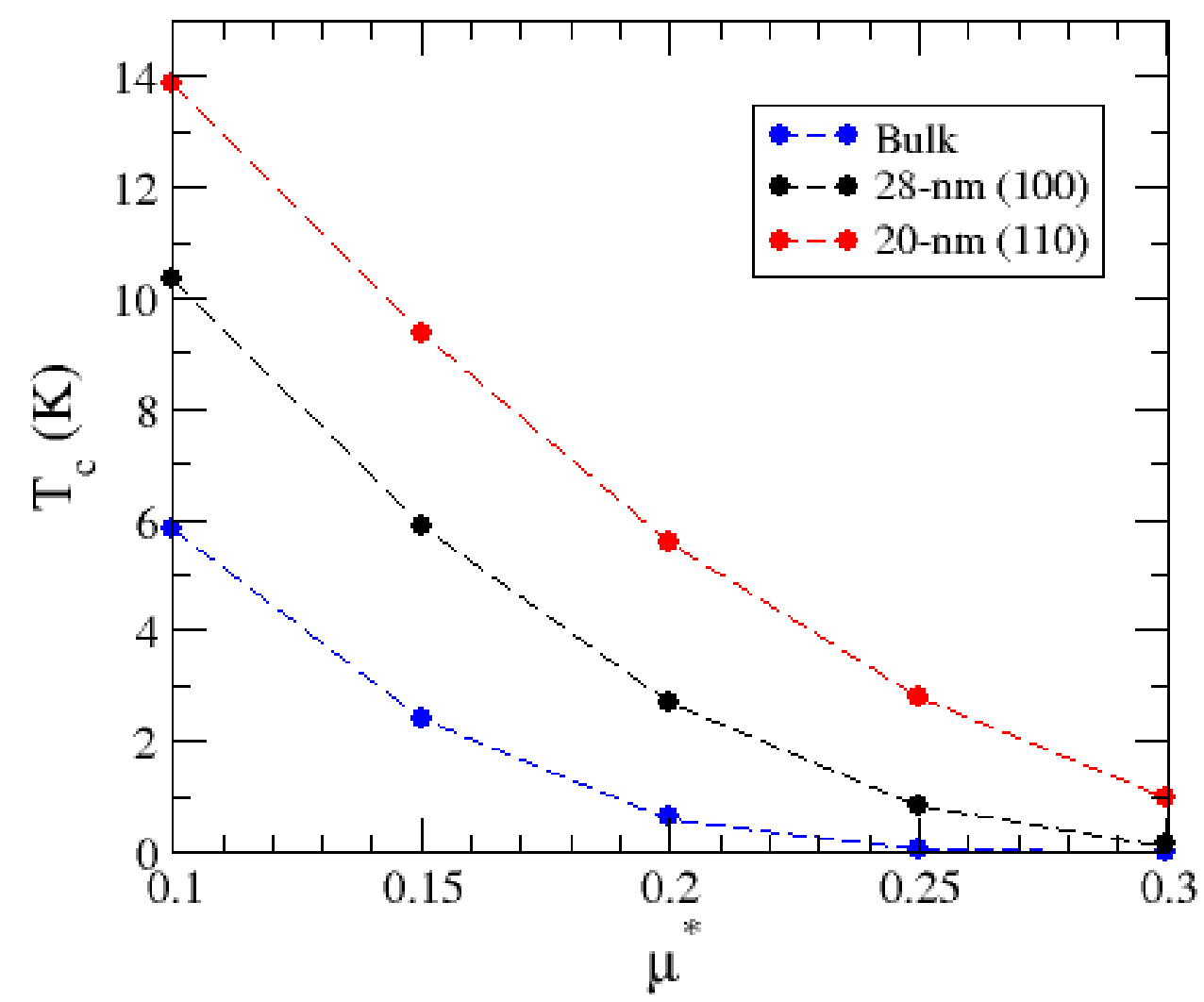

FIG. 9. Calculated superconducting transition temperature $T_c$ as a function of the Coulomb pseudopotential $\mu^*$ for bulk $RuO_2$, 28-nm $RuO_2$ (100) film, and 20-nm $RuO_2$ (110) film.

### B. On the thickness dependence of $T_c$ in ultrathin $RuO_2$ films

As presented above, the present *ab initio* calculations reproduce the overall experimental trend that superconductivity in $RuO_2$ (110) and $RuO_2$ (100) films is enhanced under epitaxial strain. Furthermore, the calculated $T_c$ of 0.83 K for the 28-nm $RuO_2$ (100) film is in good agreement with the measured value of 0.97 K [23] (Table V), and the calculated $T_c$ for 20-nm $RuO_2$ (110) film is 2.77 K, being close to the experimental value of 1.97 K (Table V). However, as mentioned before, the superconductivity is not observed in ultrathin $RuO_2$ films ($t \leq 10$ nm) (Table V) (see [18,19,23]), in contrast to the monotonical increase of the calculated $T_c$ with the decreasing thickness, which results from the larger compressive strain (Fig. 8).

Previous experimental studies suggested that this behavior of the superconductivity in ultrathin $RuO_2$ films originates from the competition between intrinsic strain-enhanced superconductivity and disorder-induced superconductivity suppression in the materials [23]. While thinner films experience larger epitaxial strains, greater enhanced EPC and hence higher $T_c$, larger disorder scattering and stronger interface-related effects become increasingly important and suppress the superconductivity [23]. The superconducting coherence lengths in $RuO_2$ films are around 15 nm [18] and superconductivity here is three-dimensional. Consequently, the superconductivity in the $RuO_2$ films with $t \leq 10$ nm tends to be significantly suppressed. Since the present strained-bulk approach deals with only the strained infinite crystals, the extrinsic effects such as disorders, out-of-plane quantum confinement and interface scattering are not included. As a result, the calculated $T_c$ would increase monotonically, independent of the film thickness, resulting in the pronounced discrepancy with the experimental findings in the ultrathin films [18,19,23].

## VII. CONCLUSION

In summary, we have investigated the effect of epitaxial strain on the superconductivity in $RuO_2$ (110) and (100) thin films both experimentally and theoretically. We have grown several $RuO_2$ (110) thin films on $TiO_2$ (110) substrates and observe superconductivity with $T_c$ of ~2.0 K in 20-nm and 30-nm $RuO_2$ (110) thin films. Theoretically, we study the superconductivity in both $RuO_2$ (110) and $RuO_2$ (100) epitaxial films based on *ab initio* calculations. First of all, we find that the calculated phonon dispersion relations of the $RuO_2$ films have no imaginary frequencies, in contrast to all previous theoretical predictions [18,19,23]. Importantly, this allows us to properly describe lattice dynamics, electron-phonon coupling and superconductivity in the considered $RuO_2$ films. Our *ab initio* calculations reveal that independent of the film orientation, the [001] compressive strain $\epsilon_{[001]}$ pushes the flat Ru $d_{z^2}$ -dominent bands towards $E_F$, and soften the out-of-plane acoustic phonon mode near the R point in the Brillouin zone, which significantly enhances the EPC strength $\lambda$ and hence the calculated $T_c$ in the $RuO_2$ films. The amount of the phonon softening and the enhanced $T_c$ increase almost linearly with the increasing $\epsilon_{[001]}$. The calculated $T_c$ for the films with thickness of ≥ 20-nm agree well with the experimental $T_c$. For example, the theoretical $T_c$ of 2.8 K for 20-nm $RuO_2$ (110) film and 0.8 K for 28-nm $RuO_2$ (100) film are close to the experimental $T_c$ of 2.0 K and 1.0 K, respectively. These findings show that lattice mismatch strain provides an effective route for tuning phonon-mediated superconductivity in $RuO_2$ and other thin films.

## ACKNOWLEDGMENTS

We thank Drs. S. C. Weng, S. Yang, C. K. Cheng, G. R. Chen, and C. H. Hsu for assistance with the X-ray measurements and analyses at NSRRC Hsinchu, Taiwan. We acknowledge the support from the National Science and Technology Council, National Center for Theoretical Sciences, and Academia Sinica, Taiwan.

---

**Supplemental Material for**

# Strain-enhanced superconductivity in $RuO_2$ thin films: Joint experimental and theoretical studies

Moeen ud din[1,2], Tsun-Sung Kuo[1], Yun-Chen Liao[1], Akhilesh Kr. Singh[3], Uddipta Kar[3], Wei-Li Lee[3,*], Guang-Yu Guo[1,4 *]

[1]Department of Physics and Center for Theoretical Physics, National Taiwan University, Taipei 10617, Taiwan, Republic of China

[2]Nano Science and Technology Program, Taiwan International Graduate Program, Academia Sinica, Taipei 11529, Taiwan, Republic of China

[3]Institute of Physics, Academia Sinica, Nankang, Taipei 11529, Taiwan, Republic of China

[4]Physics Division, National Center for Theoretical Sciences, Taipei 10617, Taiwan, Republic of China

This supplemental material contains supplemental notes A and B, figures S1, S2, and S3.

## A. Electronic band structures of $RuO_2$ films.

Figure S1 presents the orbital-projected band structures as well as total and Ru *d* and O *p* orbital-decomposed density of states (DOS) for bulk $RuO_2$, 20-nm $RuO_2$ (110) and 28-nm $RuO_2$ (100) films. Figure S1(a) indicates that in bulk $RuO_2$, four Ru *d*-orbital dominated bands cross the Fermi level ($E_F$). Notably, the doubly degenerate bands along Γ-Z crossing $E_F$ (green curves) arise mainly from Ru $d_{xz}$- and $d_{yz}$-orbitals, while other bands crossing $E_F$ (brown curves) are dominated by Ru $d_{x^2-y^2}$-orbitals. In 20-nm $RuO_2$ (110) and 28-nm $RuO_2$ (100) films, the orbital characters of the bands near $E_F$ remain more or less the same [Figs S1(c) and S1(e)]. Figure S2 shows the band structures and total DOS for 30-nm, 8-nm, and 3-nm $RuO_2$ (110) films as well as 17-nm $RuO_2$ (100) film. Figure S2(a) indicates that the band structures of 30-nm, 8-nm, and 3-nm $RuO_2$ (110) films remain qualitatively similar to that of bulk $RuO_2$, with four Ru *d*-derived bands crossing $E_F$, confirming that the metallic character is preserved over the full thickness range studied here. One pronounced change is as follows. As the compressive strain $\epsilon_{[001]}$ increases with the decreasing film thickness (see Tables II and III in the main text), Ru $d_{z^2}$-orbital dominated doubly degenerate bands below the Fermi level along the Z-R-A-Z direction (see red curves in Fig. S1) shift monotonically towards $E_F$, reflecting the increasing effect of the contracting *c* lattice constant. Nevertheless, for the considered $RuO_2$ (110) and (100) films, the Ru $d_{z^2}$-orbital dominated bands remain below $E_F$ (see Figs. S1 and S2) and hence, their upwards shift hardly leads to a significant increase in the DOS at $E_F$ [$N(E_F)$] (see Table V), as discussed in the main text.

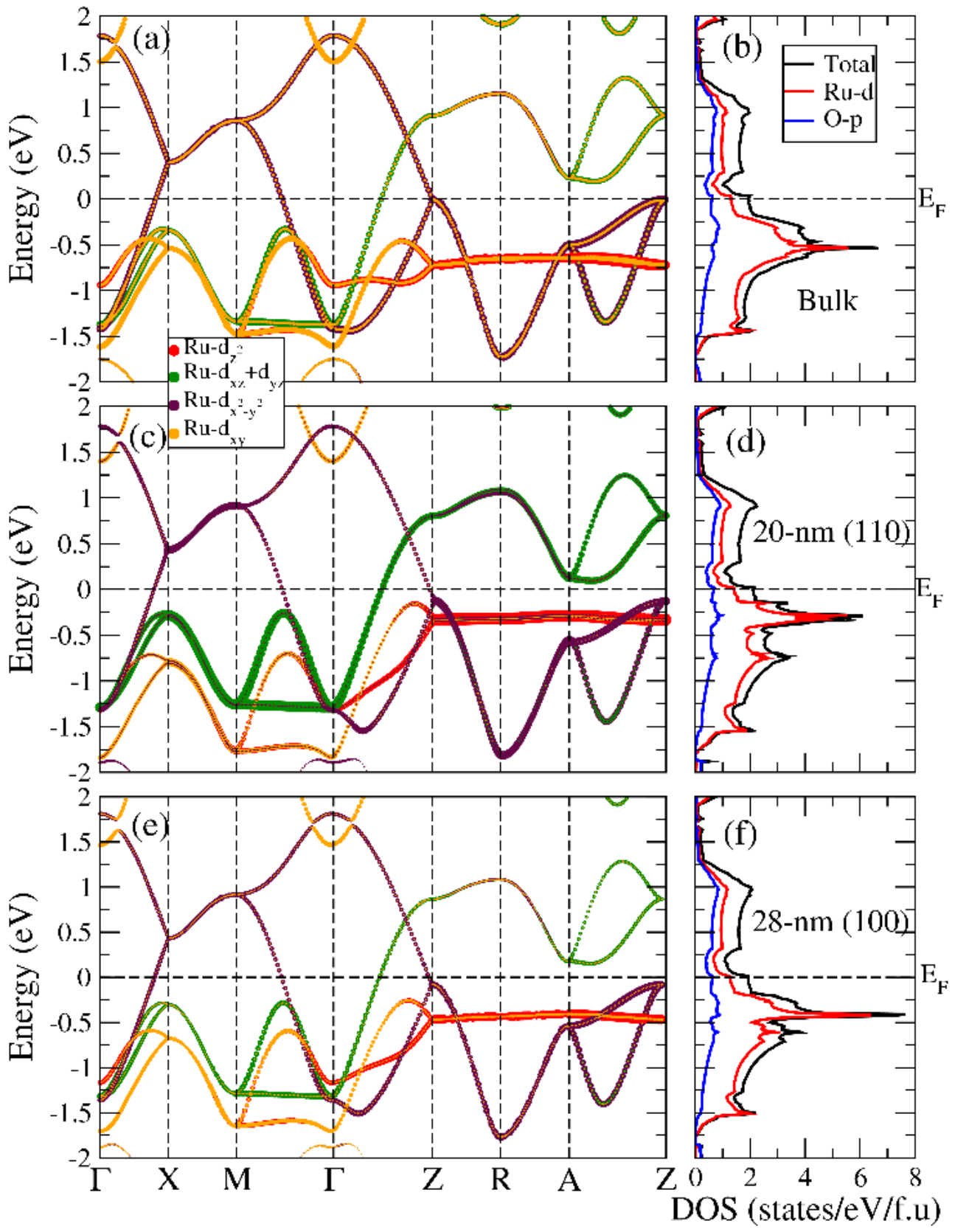

FIG.S1. (a, c, e) Orbital-projected band structures and (b, d, f) total and orbital-decomposed density of states (DOS) for bulk $RuO_2$, 20-nm $RuO_2$ (110), and 28-nm $RuO_2$ (100) films. The Fermi level is set to zero.

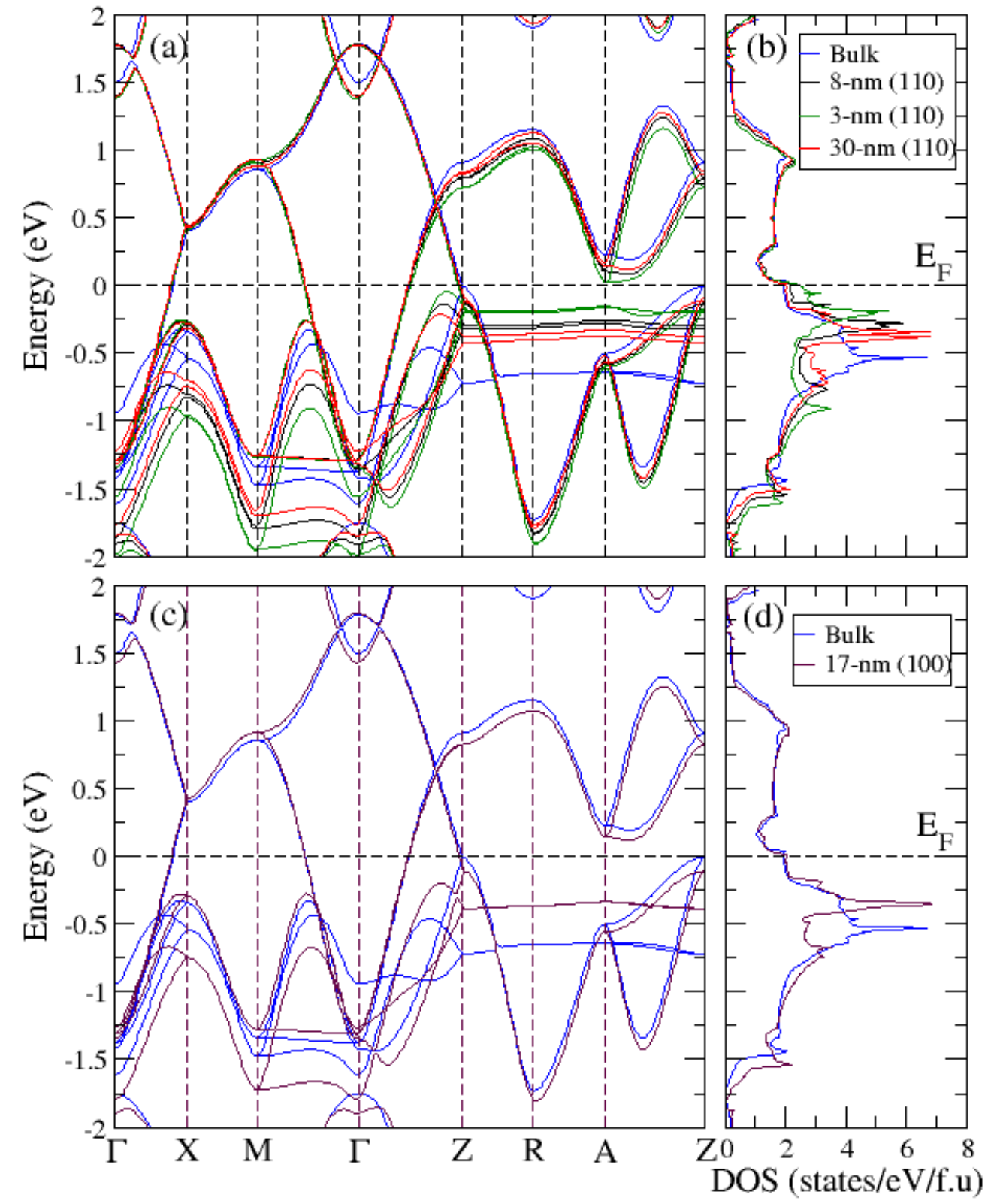

FIG.S2. (a, c) Electronic band structures and (b, d) the total density of states (DOS) for 30-nm,8-nm, and 3-nm $RuO_2$ (110) and 17-nm $RuO_2$ (100) films. The Fermi level is set to zero.

## B. Phonon dispersions and electron-phonon coupling in strained $RuO_2$ films

Figure S3 presents the phonon dispersions, phonon DOS (PhDOS), Eliashberg spectral functions $\alpha^2F(\omega)$, cumulative electron-phonon coupling strength $\lambda(\omega)$ for 30-nm, 8-nm, and 3-nm $RuO_2$ (110) films and 17-nm $RuO_2$ (100) film. The same results for 20-nm (110) and 28-nm (100) films have already been displayed in Fig. 5 in the main text. Overall, the phonon dispersions of all the considered $RuO_2$ films exhibit similar trends as bulk $RuO_2$ (see Fig. 5 and Fig. S3), as discussed in the main text. Importantly, all the calculated phonon dispersions exhibit no imaginary frequencies, in contrast to all the previous *ab initio* calculations [1-3], i.e., all the considered $RuO_2$ (110) and $RuO_2$ (100) films are dynamically stable. This allows us to evaluate the superconducting properties such as $T_c$ of $RuO_2$ films more accurately than before. One prominent change in the phonon dispersion caused by the compressive strain in strained $RuO_2$ films is the following. In all considered $RuO_2$ films, the low-energy (below 40 meV) optical and out-of-plane acoustic branches along the Γ-Z-R-A lines exhibit significant softening (see Fig. 5 and Fig. S3). Furthermore, as the compressive strain $\epsilon_{[001]}$ increases with the decreasing film thickness, this phonon softening become monotonically greater. Importantly, these softening modes give rise to a large spectral weight in the Eliashberg functions $\alpha^2F(\omega)$ in the low frequency region in the $RuO_2$ films [Figs. 5(c) and 5(f) as well as Figs. S3(c) and S3(f)], thereby resulting in a significant increase in the electron-phonon coupling constant $\lambda$ [See Eq. (3)]. In particular, Table V in the main text indicates that $\lambda$ increases from 0.5 for bulk $RuO_2$ to 0.78 for 20-nm $RuO_2$ (110) film to 0.95 for 3-nm $RuO_2$ (110) film. This leads to the monotonically enhanced $T_c$ with the decreasing film thickness due to the increasing lattice mismatch strains.

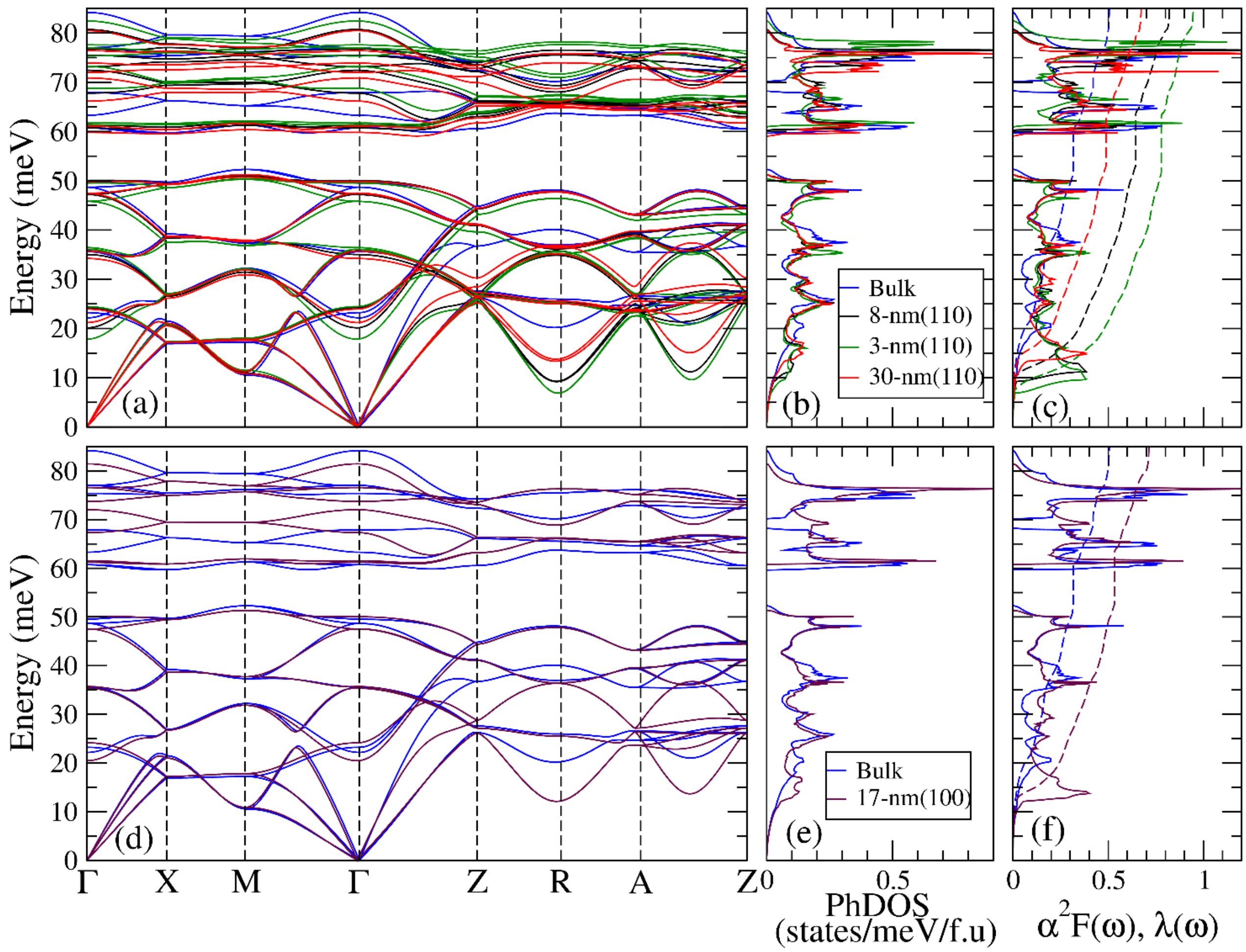

FIG. S3. (a, d) Phonon dispersions, (b, e) phonon density of states (PhDOS), (c, f) Eliashberg spectral function $\alpha^2F(\omega)$ and cumulative electron-phonon coupling strength $\lambda(\omega)$ (dashed lines) for the $RuO_2$ films.